\documentclass[rmp,notitlepage,nolongbibliography,amssymb]{revtex4-1}
\usepackage{graphicx}
\usepackage{amsmath}
\usepackage{color}
\setcitestyle{numbers,square}
\begin{document}

\def \set#1{\ifmmode \{\,#1\,\}\else $\{\,#1\,\}$\fi}
\def \q#1 {\bigskip \noindent #1.\enskip}
\def \Re{\mathop{\rm Re}\nolimits}
\def \Im{\mathop{\rm Im}\nolimits}
\def \mod{\mathop{\rm mod}\nolimits}
\def \tr{\mathop{\rm tr}\nolimits}
\def \det{\mathop{\rm det}\nolimits}
\def \b {\bigskip\noindent}
\def \m {\medskip\noindent}
\def \s {\smallskip\noindent}
\def \q#1 {\bigskip \noindent #1.\enskip}
\def \pf {\m {\bf Proof.\enskip}}
\def \sol {\m {\bf Solution.\enskip}}
\def \qed {\smallskip\rightline{QED}}
\def \pts#1 {\smallskip\rightline {(#1 pts)}}
\def \downstrut{\vrule depth4.5pt width0pt}
\def \A{ {\cal A}}
\def\pot{\Phi}
\def\ham{{\cal H}}

\title{%
Is the Ott-Antonsen manifold attracting?
}
\author{Jan R.~Engelbrecht$^{1}$ and Renato Mirollo$^{2}$}
\address{Departments of Physics$^{1}$ and \mbox{Mathematics,$^{2}$}
 Boston College, Chestnut Hill, MA 02467}
\begin{abstract}
{\bf Abstract:}
The Kuramoto model is a  paradigm for studying  oscillator networks with interplay between coupling tending towards synchronization,
and heterogeneity in the oscillator population driving away from synchrony.
In continuum versions of this model an oscillator population is represented by a probability density on the circle.
Ott and Antonsen identified a special class of densities which is invariant under the dynamics 
and on which the dynamics are low-dimensional and analytically tractable.
The reduction to the OA manifold has been used to analyze the dynamics of many variants of the Kuramoto model.
To address the fundamental question of whether the OA manifold is attracting,
we develop a systematic technique 
using weighted averages of Poisson measures
for analyzing dynamics off the OA manifold.
We show that for models with a finite number of populations, the OA manifold is {\it not} attracting in any sense;
moreover, the dynamics off the OA manifold is often more complex than on the OA manifold,
even at the level of macroscopic order parameters.
The OA manifold consists of Poisson densities $\rho_\omega$.
A simple extension of the OA manifold consists of averages of pairs of Poisson densities;   then the
hyperbolic distance between the centroids of each Poisson pair is a dynamical invariant (for each $\omega$).
These conserved quantities, defined on the double Poisson manifold, are a measure of the distance to the OA manifold. 
This invariance implies that 
chimera states, which have some but not all populations  in sync,
can never be stable in the full state space,
even if stable in the OA manifold.
More broadly,
our framework
facilitates
the analysis of multi-population
continuum Kuramoto networks beyond the restrictions of the OA manifold, 
and has the potential to reveal more intricate
dynamical behavior than has previously been observed for these networks.
\end{abstract}
\maketitle

\section{Introduction}

\noindent
The Kuramoto oscillator model, first proposed by Kuramoto in 1975 
\cite{kuramoto1975self},
is the dynamical system governed by the equations
\begin{equation}
\label{Ksys}
 \dot \theta_j = \omega_j+{K \over N} \sum_{k = 1}^N  \sin(\theta_k -  \theta_j), \quad j = 1, \dots, N. 
\end{equation}
Here $\theta_j$ is an angular variable, which we can think of as representing a point on the unit circle $S^1$, so the state space for this system is the $N$-fold torus $T^N = (S^1)^N$.   The so-called natural frequencies $\omega_j$ are typically chosen randomly according to some frequency distribution, but do not vary in any particular realization of the model.  The constant $K$ controls the nature of the system coupling; roughly speaking when $K >0$ the coupling term in 
(\ref{Ksys})\ 
tends to draw the oscillators closer to synchrony, whereas variation in the natural frequencies tends to push the oscillators away from sync.  
Over the years since its inception the Kuramoto system has become a standard paradigm to model oscillator networks with interplay between coupling tending towards synchronization and heterogeneity in the oscillator population driving away from sync.

Note that the coupling in 
(\ref{Ksys})\ 
is all-to-all, and the coupling term between any two individual oscillators is identical.  One can consider generalizations of 
(\ref{Ksys})\ 
for which the oscillators are attached to the nodes of a graph, and each oscillator is coupled only to its adjacent oscillators.  One can also introduce variation in the coupling strengths across the graph edges, so that the coupling coefficients are given by an $N \times N$ matrix.  Other variations are possible, including the introduction of phase lag terms in the couplings, by replacing $\sin (\theta_k - \theta_j)$ with $\sin (\theta_k - \theta_j - \alpha_{jk}$).  In this paper, to keep the exposition as simple as possible, we will stick with the all-to-all version 
(\ref{Ksys}), 
though our results easily generalize to the variations we described.

The system 
(\ref{Ksys})\ 
is often referred to as the finite-$N$ Kuramoto model.  In the 1990's researchers began to study continuum limit analogues of the finite-$N$ Kuramoto model, which in many ways are easier to analyze than the finite-$N$ model 
\cite{strogatz1991stability,strogatz2000kuramoto,pikovsky2008partially}.  
More recently, in their seminal paper 
\cite{ott2008low}, 
Ott and Antonsen identified a special subspace in these continuum limit systems which is invariant under the dynamics (detailed definitions will follow below). 
The asymptotic dynamics on this OA manifold can often be fully described.  
(Reference \cite{bick2019understanding} has a nice presentation of the OA technique and some of its many applications.)
This of course leads to the important question that is the title of this paper; is this OA manifold attracting for the dynamics in the full state space?  If this were the case, then the long-term dynamics on the full state space would be identical to that on the OA manifold.  To cut to the chase, we will prove below that the answer to this question is in some cases ``no,'' and in some cases ``it depends.''  The distinction between the cases is whether the continuum Kuramoto model analogue under consideration consists of a finite number $N$ of oscillator populations, each with a given natural frequency $\omega_j$, or a continuous distribution of oscillator populations with natural frequency $\omega \in \Bbb R$.  We will refer to these two versions as the finite-$N$ and infinite-$N$ continuum systems respectively.

The OA manifold is not attracting in the finite-$N$ continuum system.  We will show this by constructing a family of invariant manifolds generalizing the OA manifold, that can be arbitrarily close to the OA manifold.  We also will construct a quantity that measures the distance to the OA manifold, and that is dynamically invariant.  This implies that the OA manifold can be at best neutrally stable for the dynamics in the full state space.  In fact, to drive home this point, we will show that the stable long-term dynamics off the OA manifold is usually  qualitatively distinct and more complex than that on the OA manifold.  
This has important ramifications for the study of finite population continuum Kuramoto models and especially the existence and stability of chimera states  
\cite{abrams2004chimera,
abrams2008solvable,
laing2009chimera,
martens2009exact,
martens2013chimera}.
Chimeras are fixed states for which some of the populations are completely synchronized, whereas others are smoothly distributed in phase.
Chimera states exist in systems with as few as $N=2$ populations, and may be stable within the OA manifold  
\cite{abrams2008solvable}.
Our analysis implies that these states are {\it never} stable in the full system state space;  moreover, the dynamics near these states but off the OA manifold are 
more complicated:
typically the steady state dynamics near chimera states but off the OA manifold will be stable limit cycles.
Our results for finite population systems also have consequences for numerical simulations:
since chimera states are only neutrally stable, 
simulations of chimera states will typically drift off the OA manifold on which they are located, unless one explicitly designs the numerical algorithm to force trajectories to remain in the OA manifold. 

We give a similar construction of families of invariant manifolds off the OA manifold in the infinite--$N$ continuum case,
as well as a measure of the distance to the OA manifold.  The dynamics on these families are also given explicitly.
However in the infinite--$N$ case we run into subtle issues concerning the topology of the full state space.  There is a natural strong topology on this state space which is derived from the natural topology on the space of probability measures on the circle.  Our constructions imply that the OA manifold is not attracting in this strong topology. 
Moreover, as in the finite--$N$ case, the steady-state dynamics of the individual oscillator populations is typically more complex off the OA manifold.
However, in a sufficiently weak topology the OA manifold can be attracting, and the dynamics of the macroscopic system order parameter may not change when we perturb off the OA manifold. 
Various versions of this result have been proved by Chiba 
\cite{chiba2011center,chiba2015proof}
and by Ott and Antonsen 
\cite{ott2009long}.
In our framework we can see this explicitly;
using techniques from hyperbolic geometry, 
we prove that the macroscopic order parameter on our extended  OA manifolds must have the same asymptotic dynamics as on the OA manifold.

\section{Finite--N Continuum System}

\subsection{System Set-up}

\noindent
We can construct a continuum version of 
(\ref{Ksys})\ 
by replacing a single oscillator $\theta_j$ with natural frequency $\omega_j$ by an infinite population of oscillators with this frequency.  This population will be represented by a probability measure $\rho_j$ on the circle $S^1$.  The space $Pr(S^1)$ of Borel probability measures on the circle has a natural topology, which is metric and compact; this is the topology induced from the inclusion in the space $C^1(S^1)^\ast$, the dual of the Banach space $C^1(S^1)$ of continuously differentiable functions on $S^1$ (see 
\cite{mirollo2007spectrum}
for details on this natural topology).  Therefore the state space for the finite-$N$ continuum system is $X = Pr(S^1)^N$. 

The finite-$N$ system 
(\ref{Ksys})\ 
in complex form with $\zeta_j = e^{i \theta_j}$ is
\begin{equation}
\label{Ksyscx}
\dot \zeta_j = i \omega_j \zeta_j + {K \over 2} \left(Z - \overline Z \zeta_j^2 \right), 
\quad {\rm where} \quad 
Z = {1 \over N} \sum_{j=1}^N \zeta_j 
\end{equation}
is the system's complex order parameter (see 
\cite{chen2017hyperbolic}
for the derivation).
In the finite--$N$ continuum system, the measures $\rho_j$ evolve according to the continuity equations
\begin{equation}
 \label{KsysFC}
\dot \rho_j + {\partial \over \partial \zeta} \left( v_j \rho_j \right) = 0,
\quad {\rm with} \quad
v_j(\zeta) =  i \omega_j \zeta + {K \over 2} \left(Z - \overline Z \zeta^2 \right),
\end{equation}
$j=1,\ldots,N$
and  order parameter
\begin{equation}
\label{ZF}
Z = {1 \over N} \sum_{j=1}^N \int_{S^1} \zeta \, d\rho_j(\zeta). 
\end{equation}
Technically, this means that $\dot\rho_j$ is the distribution on $S^1$ defined by
\begin{equation}
\label{rhodot}
\langle f, \dot \rho_j \rangle = \int_{S^1} f'(\zeta) v_j(\zeta) \, d \rho(\zeta), 
\end{equation}
for smooth $f$ on $S^1$.  We note that if we take each $\rho_j$ to be a unit mass measure (delta function) at some point $\zeta_j \in S^1$, then the system 
(\ref{KsysFC})\ 
reduces to 
(\ref{Ksyscx}). 

Let $G$ be the 3D M\"obius group consisting of the M\"obius transformations which preserve the unit disc $\Delta$, and therefore also the boundary circle $S^1$.  These transformations have the form
$$
M(z) = \zeta {z - z_0 \over 1-\overline z_0 z},
$$
where the parameters satisfy $|z_0| < 1 $ and $|\zeta| = 1$. The group $G$ plays an important role in complex analysis and hyperbolic geometry:  $G$ is the group of orientation-preserving isometries of the disc with the hyperbolic metric given by
$$
ds = {2 |dz| \over 1 - |z|^2}.
$$
The hyperbolic metric is calculated as follows 
\cite{ahlfors1973conformal}:  
for any $z, w \in \Delta$, define
\begin{equation}
\label{HYPONE}
\lambda (z,w) = {z-w \over 1- z\overline w } \in \Delta \quad { \rm and} \quad \delta(z,w) = | \lambda(z,w)|. 
\end{equation}
Then $\delta(z,w)$ is invariant under $G$, and the hyperbolic distance is given by
\begin{equation}
\label{HYPTWO}
d_{hyp} (z,w) = \log {1+\delta \over 1-\delta} = 2 \left( \delta + {\delta^3 \over 3} + {\delta^5 \over 5} + \cdots \right). 
\end{equation}
Notice that for $z, w \approx 0$ this gives $d_{hyp} (z,w) \approx 2 |z-w|$ as expected, since the hyperbolic metric $ds \approx 2 |dz|$ near $0$.

The action $M_\ast \rho$ of an element $M \in G$ on a measure $\rho \in Pr(S^1)$ is determined by the adjunction formula
\begin{equation}
\label{Maction}
\int_{S^1} f(\zeta) \, d (M_\ast \rho) (\zeta) = \int_{S^1} f(M(\zeta) )\, d\rho (\zeta), 
\end{equation}
where $f$ is any continuous function on $S^1$.  There is a natural action of the group $G^N$ on the state space $X$: the action of an element $(M_1,  \dots, M_N) \in G^N$ on a state $(\rho_1, \dots, \rho_N) \in X$ is just given coordinate-by-coordinate using the $G$-action above.  

The infinitesimal generators for the action of $G$ on the circle are the vector fields of the form
$$
v(\zeta) = i \omega \zeta  + Z - \overline Z \zeta^2,
$$
with $\omega \in \Bbb R$ and $Z \in \Bbb C$ constants (this is derived in 
\cite{marvel2009identical}).  
Since the function $v_j(\zeta)$ in 
(\ref{KsysFC})\ 
has this form, the dynamical trajectory of a density $\rho_j$ under 
(\ref{KsysFC})\ 
must remain in its group orbit $G \rho_j$, and so the dynamical trajectories of 
(\ref{KsysFC})\ 
are constrained to lie on $G^N$ group orbits.  This implies that the infinite-dimensional system 
(\ref{KsysFC})\ 
can be reduced to a finite-dimensional system with dimension at most $3N$. 
The group orbits are  typically 3D, except in an important special case:  the orbit of the uniform density $m$ consists of the 2D space of Poisson measures.

\subsection{Poisson Manifold $X_P$}

\noindent
Poisson measures arise naturally in complex analysis in the solution to the Dirichlet problem on the unit disk.
If $u(\zeta)$ is continuous on the unit circle $S^1$, then $u$ has a unique continuous extension to a harmonic function $\tilde u$ on the disk,
which can be expressed as follows.  For any $z\in\Delta$, 
$$
\tilde u(z)=\int_{S^1} u(\zeta) \; d \rho(\zeta),
$$
where
 $\rho$ is the  Poisson measure on $S^1$
 given by
$$
\begin{aligned}
d\rho(\zeta) &= {1 \over 2\pi i} \left( 1 + \sum_{n=1}^\infty (\overline z \zeta)^n + \sum_{n=1}^\infty ( z \overline \zeta)^n\right)  {d \zeta \over \zeta}
\\&
= {1 \over 2\pi i} \left(   { 1 \over \zeta-z }+  {\overline z  \over 1- \overline z \zeta} \right)   d \zeta  \cr
&=  { 1 \over 2 \pi i} {1 - |z|^2 \over |\zeta - z|^2} \cdot  {d \zeta \over \zeta}
.
\end{aligned}
$$
The measure $\rho$ has  centroid
$$
\int_{S^1} \zeta \, d \rho(\zeta)=z;
$$
more generally, 
the Cauchy integral formula shows that the moments of $\rho$ are
\begin{equation}
\label{mom}
 \int_{S^1} \zeta^n \, d \rho(\zeta)= z^n, \quad n \ge 0  . 
\end{equation}
Poisson measures are invariant under the group $G$; in fact, the Poisson measures  are precisely the group orbit of the uniform measure (which is the Poisson measure with centroid $z = 0$).  To see this, we express the uniform measure $m$ on $S^1$ in the form
$$
dm(\zeta) = {1 \over 2 \pi i} {d\zeta \over \zeta}
$$
and let $M \in G$ be any M\"obius map fixing $\Delta$.  Then for all continuous $f$ on $S^1$,
$$
\int_{S^1} f(\zeta) d(M_\ast m) (\zeta) = {1 \over 2 \pi i} \int_{S^1} f(M(\zeta)) {d \zeta \over \zeta} = {1 \over 2 \pi i} \int_{S^1} f(\zeta) {d M^{-1} (\zeta) \over M^{-1}(\zeta)}.
$$
Suppose $M(0) = z$; then $M^{-1}(z) = 0$, and we can express $M^{-1}$ in the form
$$
M^{-1} (\zeta) = \alpha {\zeta - z \over 1 - \overline z \zeta}
$$
with $|\alpha| =1$.  Therefore
$$
d(M_\ast m)(\zeta) 
=  {1 \over 2 \pi i} {d M^{-1} (\zeta) \over M^{-1}(\zeta)} 
= {1 \over 2 \pi i} \left ( {1 \over \zeta - z} + {\overline z \over 1- \overline z \zeta} \right) d \zeta, 
$$
so $M_\ast m $ is the Poisson measure with centroid $z = M(0)$.  We also include delta functions among the Poisson measures, since they arise as limits of smooth Poisson measures, so the complete space of Poisson measures is equivalent to the closed unit disc $\overline \Delta$, parametrized by the centroid $z \in \overline \Delta$.  {\sl For the finite-$N$ continuum system, the OA manifold is the Poisson manifold $X_P$ consisting of $N$-tuples of Poisson measures.}  Topologically, $X_P$ is $\overline \Delta^N$, a $2N$-dimensional compact manifold with boundary.

The dynamics on the Poisson manifold in terms of the centroids $z_j$ of the Poisson measures $\rho_j$ can be derived as follows:  let $f(\zeta) = \zeta$ and use 
(\ref{rhodot})\ 
and 
(\ref{mom}): 
$$
\begin{aligned}
\dot z_j &= \langle f, \dot \rho_j \rangle = \int_{S^1} 1 \cdot \left ( i \omega_j \zeta + {K \over 2} \left(Z - \overline Z \zeta^2 \right) \right) d \rho_j( \zeta) 
\\&
= i \omega_j z_j + {K \over 2} \left( Z - \overline Z z_j^2 \right).
\end{aligned}
$$
So the dynamics on the Poisson manifold are given by the system
\begin{equation}
\label{PsysFC}
\dot z_j  =  i \omega_j z_j + {K \over 2} \left( Z - \overline Z z_j^2 \right) \quad
 {\rm with} \quad
Z = {1 \over N} \sum_{j=1}^N z_j. 
\end{equation}
Not coincidentally, this extends the equations for the finite-$N$ system 
(\ref{Ksyscx}). 
So we can think of the system on the Poisson manifold $X_P$ as an extension of 
(\ref{Ksyscx}),
where now the variables $z_j$ are no longer constrained to lie on the circle, and instead can also lie in the unit disc $\Delta$.

The system 
(\ref{PsysFC})\ 
has a fixed point with all $z_j = 0$, which corresponds to the ``incoherent state'' with all $N$ oscillator populations uniformly distributed on the circle.  The linearization of 
(\ref{PsysFC})\ 
at this state is the system
$$
\dot z_j  =   i \omega_j z_j + {K \over 2}Z, \quad j = 1, \dots, N. 
$$
An eigenvalue $\lambda$ for this system corresponds to a nontrivial solution to the linear system
$$
(\lambda - i \omega_j) z_j  =  {K \over 2}Z, \quad j = 1, \dots, N. 
$$
Let us assume for simplicity that the $\omega_j$ are distinct, and $K \ne 0$.  Then it is easy to see that any nontrivial solution must have $Z \ne 0$; otherwise, exactly one $z_j \ne 0$, but this implies $Z \ne 0$.  Hence we can assume WLOG that $Z = 1$, and therefore
$$
z_j = {K \over 2} (\lambda - i \omega_j)^{-1},
$$
so the eigenvalues $\lambda$ must satisfy  the self-consistency equation
$$
{1 \over N} \sum_{j=1}^N (\lambda - i \omega_j)^{-1} = {2 \over K}.
$$
Observe that $\Re \lambda \ge 0 \Rightarrow \Re(\lambda - i \omega_j)^{-1} \ge 0$; hence when $K <0$, all the eigenvalues satisfy $\Re \lambda < 0$ and so the incoherent state is attracting in the Poisson manifold.

\subsection{Multi-Poisson Manifolds: Dynamics off $X_P$}

\noindent
We can embed the Poisson manifold in a larger invariant manifold by considering probability measures $\rho_j$ which are averages of two Poisson measures with possibly distinct centroids:
$$
\rho_j = {1 \over 2} \left( \rho_j^{(1)} + \rho_j^{(2)}\right),
$$
which is uniquely determined by a pair of centroids $(z_j^{(1)}, z_j^{(2)})$.  The dynamical evolution of any ``double Poisson'' measure $\rho_j$ is given by $M_\ast$ for some $M \in G$, and we see from 
(\ref{Maction})\ 
that $M_\ast$ acts linearly on measures on $S^1$.  Therefore the manifold $X_{DP}$ consisting of 
double Poisson measures is invariant under the dynamics, and contains the Poisson manifold as the submanifold given by  $z_j^{(1)} =  z_j^{(2)}$.  Topologically, $X_{DP}$ is $\overline \Delta^{2N}$, a $4N$-dimensional compact manifold with boundary. The dynamics on the double Poisson manifold are given by the equations
\begin{equation}
\label{DPsysFC}
\begin{aligned}
\dot z_j^{(\alpha)} &=i \omega_j z_j^{(\alpha)} + {K \over 2} \left( Z - \overline Z (z_j^{(\alpha)})^2 \right), 
\\
Z &= {1 \over 2N} \sum_{j=1}^N (z_j^{(1)} +z_j^{(2)}) 
\end{aligned}
\end{equation}
with
$j = 1, \dots, N, \quad \alpha = 1,2.$
Note that the system 
(\ref{DPsysFC})\ 
is equivalent to the system 
(\ref{PsysFC})\ 
with $2N$ populations, and the natural frequencies occurring in pairs.

Take any point in the interior of $X_{DP}$, so all $|z^{(\alpha)}_j| < 1$ (i.e.~no delta function components). The evolution equation in 
(\ref{DPsysFC})\ 
for each pair $(z_j^{(1)}, z_j^{(2)})$ is an infinitesimal generator for the M\"obius group action on the unit disc $\Delta$, and therefore is an infinitesimal isometry for the hyperbolic metric on $\Delta$.  This means that the hyperbolic distances
$$
\ell_j = d_{hyp} ( z_j^{(1)} ,z_j^{(2)})
$$
are invariant under the dynamics given by 
(\ref{DPsysFC}). 
The vector
$$
\vec \ell = (\ell_1, \dots, \ell_N)
$$
effectively measures the distance of a point $(\rho_1, \dots, \rho_N)$ in $X_{DP}$ to the Poisson manifold $X_P$, and $\vec\ell$ is invariant under the dynamics.  We could also define a scalar invariant $\ell \ge 0$ by taking the norm of $\vec \ell$ or the average of the $\ell_j$.  This invariant $\vec \ell$ is defined on the interior of $X_{DP}$ but undefined on the boundary of $X_{DP}$  since the hyperbolic distance between two distinct points on the circle is infinite.

{\sl The invariance of $\vec \ell$ implies that the Poisson manifold is not attracting}: the trajectory of any initial condition in the interior of $X_{DP}$ with at least one $\ell_j > 0$ cannot converge to an interior point in $X_P$.  In particular, the incoherent state with all $z_j = 0$, which is attracting in $X_P$ for $K < 0$,  {\sl is not attracting} in the larger manifold $X_{DP}$.  Note that a trajectory starting in $X_{DP} - X_P$ can converge to point on the boundary of $X_P$.
For example, 
if $K > 0$ and all $\omega_j$ are equal,
then 
the 1D synchronous manifold,
which has  all $z^{(\alpha)}_j\in S^1$ and all $z^{(\alpha)}_j$
equal,
is attracting in 
$X_{DP}$
\cite{chen2019dynamics}.

More generally, we can construct invariant manifolds consisting of finite weighted averages of Poisson measures, with any finite set of weights $w_\alpha >0$ summing to 1;
these generalized Poisson manifolds are parametrized by centroids $z_j^{(\alpha)}$.	 
As in the double Poisson case, the hyperbolic distances $d_{hyp}(z_j^{(\alpha)},z_j^{(\beta)})$ are  invariant under (\ref{KsysFC}). 
For the dynamical invariant $\vec\ell$, we can let 
$\ell_j=\max d_{hyp}(z_j^{(\alpha)},z_j^{(\beta)})$.
At the end of this section we prove that these
generalized Poisson manifolds are in fact dense in the full state space $X$. 

\begin{figure}[h]
	\centerline{\includegraphics[width=2.8in]{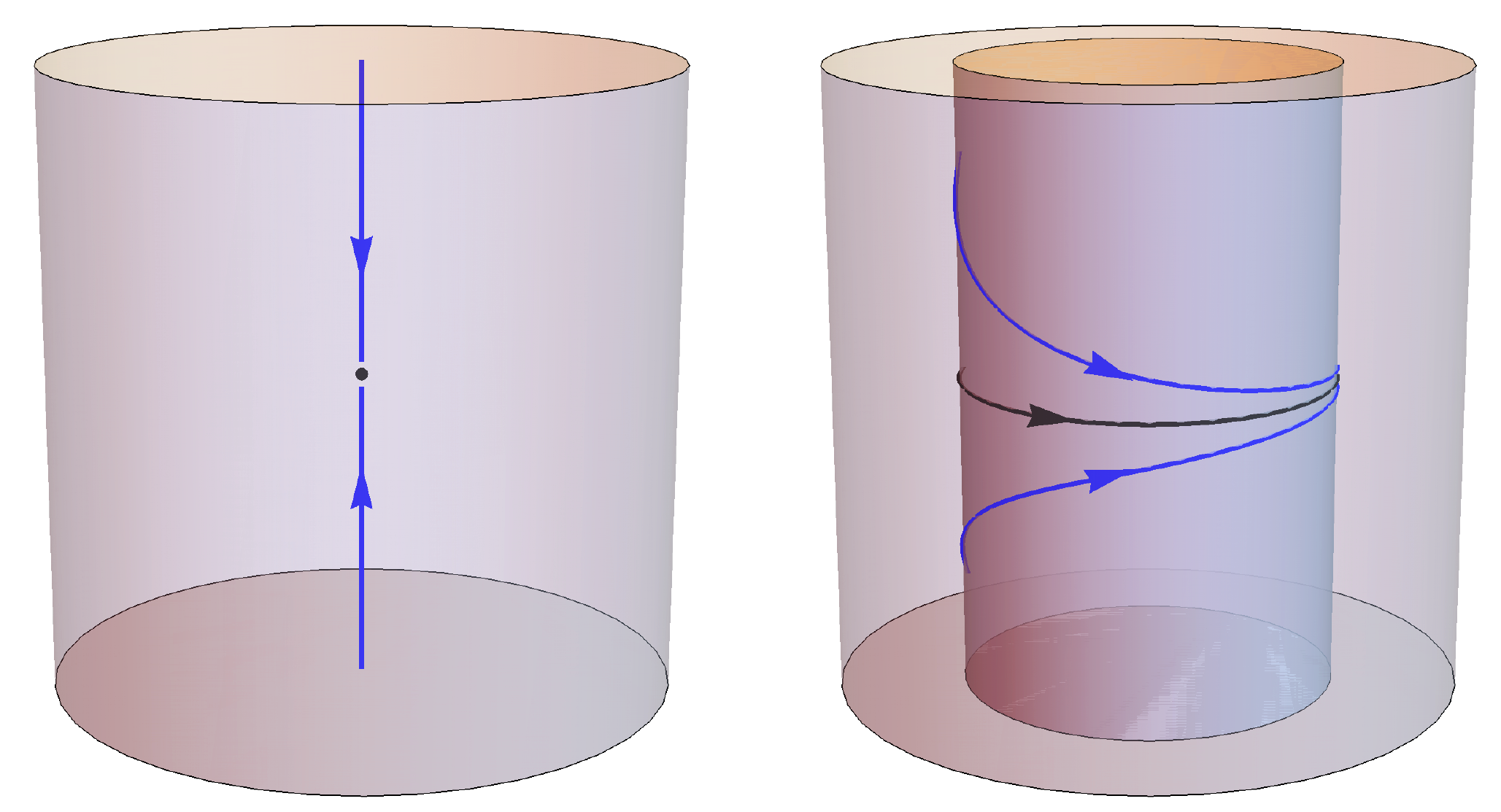}}
	\caption{\label{fig1}
		Left: 1D flow on $C_0$  toward fixed point.
		Right: 2D flow  on $C_r$  toward limit cycle.
	}
\end{figure}

So we see that the dynamics on the Poisson manifold $X_P$ are not in general attracting; they are also deceptively simple compared to the dynamics off $X_P$.  This is because $X_P$ has dimension $2N$, whereas a typical M\"obius group orbit off $X_P$ will have dimension $3N$.  We can construct a simple model in ${\Bbb R}^3$ that illustrates this phenomenon (see Figure~1).  
Consider the linear system
$$
\begin{aligned}
\dot x &= -y \cr
\dot y &= x \cr
\dot z &= -z.
\end{aligned}
$$
The group $G = S^1 \times \Bbb R$, consisting of rotations around the $z$-axis together with translations $z \to z+c$, acts on ${\Bbb R}^3$, and the group orbits are the cylinders $C_r$ of radius $r \ge 0$ centered around the $z$-axis; $C_0$ is the $z$-axis itself.  Clearly, the trajectories of the linear system above are constrained to stay in the group orbits.  The special orbit $C_0$ is analogous to the Poisson manifold; within this 1D group orbit, the origin is an attracting fixed point. 
If we move to a nearby group orbit $C_r$ with $r > 0$, then all trajectories 
on $C_r$
converge to the 
periodic orbit
consisting of the circle with $z = 0$ in $C_r$
which is a limit cycle for the dynamics restricted to $C_r$. 
Due to the collapse of one dimension at the special group orbit $C_0$, the dynamics on $C_0$ do not capture the more complicated stable steady-state dynamics on $C_r$ for $r > 0$.

\begin{figure}[h]
	\centerline{\includegraphics[height=1.7in]{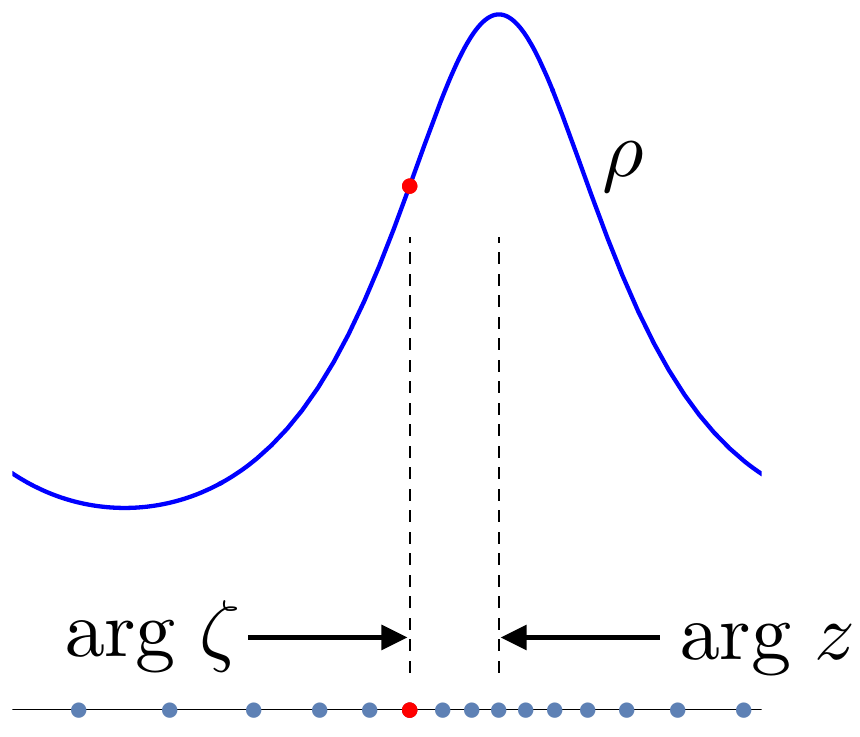}}
	\caption{\label{fig2}
		In the dynamics of a marked Poisson density ($z,\zeta$) the phase difference between the mark and the density peak typically vary in time, effectively adding an additional dimension to the dynamics.
	}
\end{figure}

Something similar happens with the loss of dimensions at the Poisson manifold, but we can get around this problem with the following trick.  Consider the augmented finite-$N$ continuum system on the state space $\tilde X$ consisting of $N$-tuples of ``marked'' densities $\tilde \rho_j = (\rho_j, \zeta_j)$; each density now has a single distinguished point or marking (see Figure~2).
We can think of a
marking as a distinguished representative oscillator among the continuum
of oscillators in the population described by the density,  which we track  as the density evolves 
(somewhat like watching the motion of a particle suspended in a fluid).	
The densities $\rho_j$ evolve according to the same equations as in 
(\ref{KsysFC}),
and the points $\zeta_j$ follow the dynamics given by the original system, with order parameter $Z$ determined by the $\rho_j$:
$$
\dot \zeta =i \omega_j \zeta_j + {K \over 2} \left (Z - \overline Z \zeta_j^2 \right), \quad j = 1, \dots, N.
$$
The points $\zeta_j$ do not contribute to the order parameter $Z$; they just ``go along for the ride'' as the densities $\rho_j$ evolve.  The $N$-fold M\"obius group $G^N$ acts on $\tilde X$ as before, and dynamical trajectories are constrained to lie inside the group orbits.  In this augmented system, the Poisson manifold $\tilde X_P$ consisting of $N$-tuples of marked Poisson densities has dimension $3N$, since for each Poisson density we can choose the marking to be any point on the circle.  Given any $\zeta, \zeta' \in S^1$ and $z, z' \in \Delta$, there exists a M\"obius transformation $M \in G$ such that $M\zeta = \zeta'$ and $Mz = z'$; hence the augmented Poisson manifold $\tilde X_P$ is a group orbit for $G^N$.  And in the augmented system $\tilde X_P$  has the same dimension $3N$ as nearby group orbits, so the dynamics on these nearby group orbits must be a continuous deformation of the dynamics on the augmented Poisson manifold.

In the augmented system, incoherent states (with all $z_j = 0$) are invariant, and the markings evolve according to the simple equations $\dot \zeta_j = i \omega_j \zeta_j$.  So the attracting steady state dynamics on the augmented Poisson manifold $\tilde X_P$ consist of periodic or quasi-periodic dynamics on the $N$-fold torus with coordinates $\zeta_j$, depending on whether the $\omega_j$ are rationally independent.  This dynamic behavior is more complicated than an attracting fixed point.  If we move to nearby group orbits in the augmented double Poisson manifold $\tilde X_{DP}$, we now can expect similar stable steady-state dynamics:  periodic or quasi-periodic dynamics on an attracting $N$-dimensional torus.  We can in fact identify these attracting tori explicitly: given any point in $\tilde X_{DP}$, we can find a point in its group orbit which has $z_j^{(1)} =  -z_j^{(2)}$ for all $j$.  Any such point has $Z = 0$, and the set of such configurations is invariant under the dynamics.  Within each group orbit, the dynamics on this invariant $N$-dimensional torus are given by $\dot z_j^{(\alpha)} = i \omega_j z_j^{(\alpha)}$, which is periodic or quasi-periodic dynamics on an invariant torus (see Figure~3).

\begin{figure}[h]
	\centerline{\includegraphics[width=2.8in]{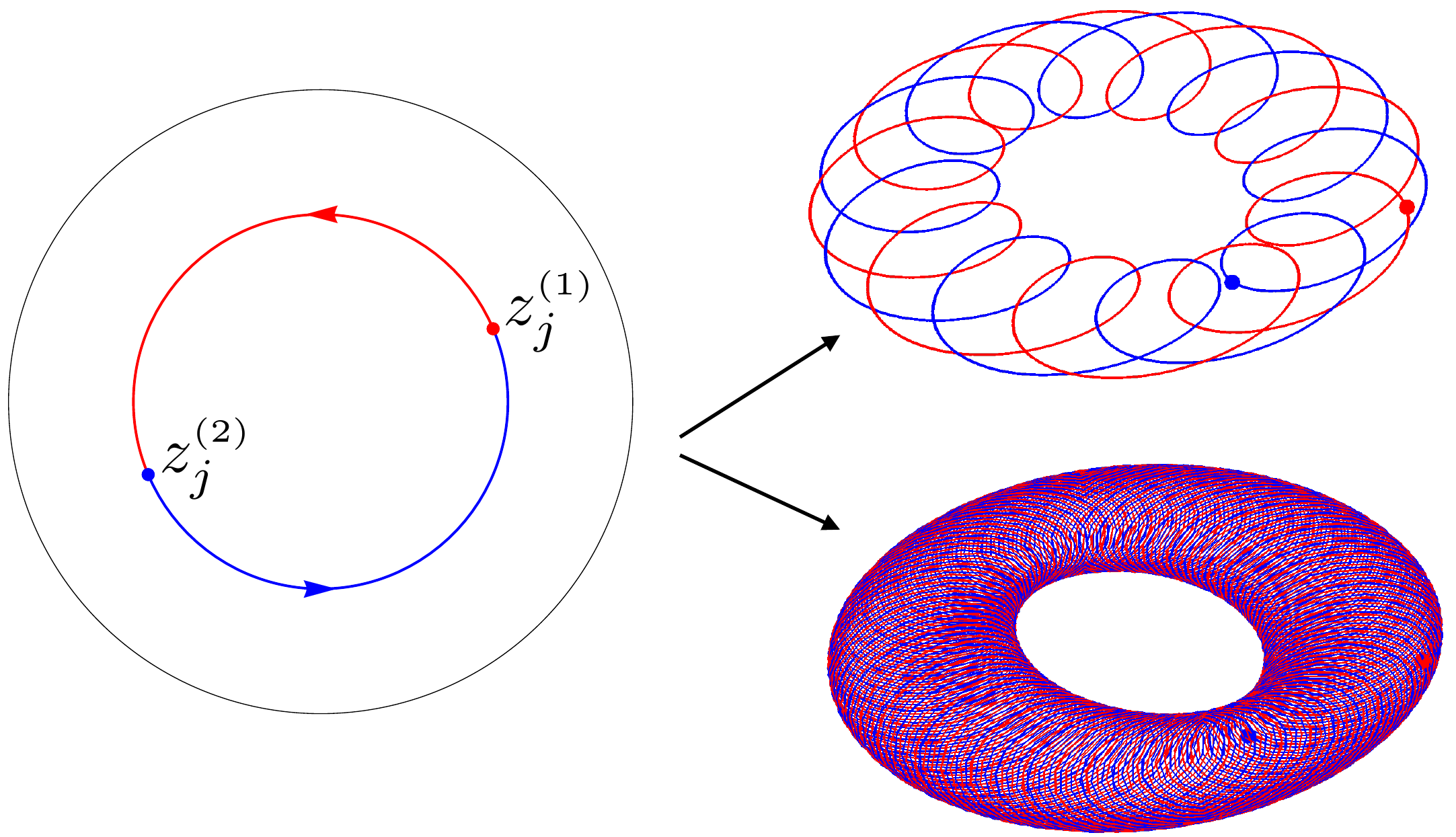}}
	\caption{\label{fig2}
		Trajectories of an antipodal pair $(z_j^{(1)}(t),z_j^{(2)}(t))$ on $\tilde X_{DP}$ for the incoherent state $Z=0$  on left; plotted relative to phase of $z_k^{(1)}$ on  right.  The ratio $\omega_j/\omega_k$ is rational for top  right (periodic) and irrational for  bottom right (quasiperiodic).	}
\end{figure}

\subsection{Order Parameter Dynamics:  An Example}

\noindent
In the example above, the steady-state dynamics off the Poisson manifold are qualitatively different from those on the Poisson manifold; however, the steady-state behavior of the order parameter $Z(t)$ is the same:  in both cases, $Z(t) \to 0$ as $t \to \infty$ for initial conditions sufficiently close to the incoherent state or nearby invariant tori. 
In this section we present an example where the dynamics off the Poisson manifold are qualitatively different than the
dynamics on it,
{\it even at the level of the order parameters}.
In this example,  we have an attracting fixed point  with order parameter $1/3$ within 
$X_P$, so nearby trajectories on $X_P$ have order parameter $Z_P(t)$ converging exponentially to $1/3$;
whereas the order parameter $Z_{DP}(t)$ for all trajectories in $X_{DP}-X_P$ {\it never} converges. 
For the sake of completeness we present a careful derivation of these assertions in the remainder of this section
(these details are not required to proceed to the subsequent sections).

Consider the system with $N=3$, index $j = -1, 0, 1$ and corresponding $\omega_j = -1, 0, 1$.  We will assume that the $j=0$ oscillator is a point $\zeta$ on the unit circle, and the $j= \pm 1$ populations are represented by Poisson measures with centroids $z_{\pm 1}$ in the disc (the assumption $|\zeta| = 1$ is necessary to find a stable fixed point; if we allow perturbations of $\zeta$ off the unit circle, we always get at least one unstable direction).  This 5D system has equations
\begin{equation}
\label{Threesys}
\begin{aligned}
\dot z_{-1} &= -iz_{-1} + {K \over 2} \left( Z - \overline Z z_{-1}^2 \right) \cr
\dot \zeta &=  {K \over 2} \left( Z - \overline Z \zeta ^2 \right) \cr
\dot z_1 &= \hphantom{-}iz_1 + {K \over 2} \left( Z - \overline Z z_1^2\right ).
\end{aligned}
\end{equation}
We look for a fixed point of the form $(-ib,  1, ib)$; this point has $Z = 1/3$, and is stationary provided that
$$
-b + {K \over 6} \left( 1+b^2\right) = 0  \iff K = {6b \over 1+b^2}.
$$
Thus we can choose any $b \in (-1,1)$ and get a fixed point for 
(\ref{Threesys})\
with $K$ given by the equation above.

We can reduce the dynamics to dimension $4$ by introducing the variables $y_{\pm 1} = \overline \zeta z_{\pm 1}$; then $\zeta$ drops out of the $y_{\pm 1}$ equations and we get the reduced system
\begin{equation}
\label{ReducedThreesys}
\begin{aligned}
\dot y_{-1} &=- iy_{-1} + {K \over 2} \left( (\overline Y - Y) y_{-1} +Y - \overline Y y_{-1}^2\right ) \cr
 \dot y_1 &= \hphantom{-} iy_1 + {K \over 2} \left( (\overline Y - Y) y_1 +Y - \overline Y y_1^2\right )
\end{aligned}
\end{equation}
with
$$
Y = \overline \zeta Z = {1 \over 3} \left ( 1 + y_1 + y_{-1}\right).
$$
Observe that this system is invariant under the involution $(y_{-1},y_1) \mapsto ( \overline y_{1}, \overline y_{-1})$.  We linearize at the fixed point $y_{\pm 1} = \pm ib$ by setting
$$
y_1 = ib + \eta, \quad y_{-1} = -ib + \nu.
$$
The linearized system is invariant under the involution $(\eta, \nu) \mapsto (\overline \nu, \overline \eta)$, and this implies that the eigenspaces of the involution, which consist of pairs $(\eta, \overline \eta)$ and $(\eta, -\overline \eta)$ respectively, are invariant under the linearized system.  Thus we can study the 2D linear systems on these eigenspaces separately.  

The linearized equation for $\eta$ is
$$
\dot \eta  = \left(
{K\over 6}\!+\!i\left(1\!-\!{Kb\over2}\right)
\right)
\eta
+{K \over 6} \left(1\!-\!ib \right) \nu
+{K \over 6} \left( b^2\!+\!ib \right) \left( \overline \eta \!+\! \overline \nu\right).
$$
On the eigenspace with $\nu = \overline \eta$ we get
$$
\begin{aligned}
\dot \eta  &= \left(  {K\over 6} \left(1\!+\!b^2\right)+i\left( 1\! -\! {Kb \over 3}  \right) \right ) \eta + {K \over 6} \left(1\! +\! b^2 \right)\overline \eta \cr
&= \left( b +\Omega i  \right ) \eta + b \overline \eta,
\end{aligned}
$$
with
$$
\Omega =  {1-b^2 \over 1+b^2}.
$$
The matrix for this 2D system with respect to the real coordinates $(\Re \eta, \Im \eta)$ is
$$
M_1 = 
\begin{pmatrix}b & - \Omega  \cr  \Omega  &\hphantom{-} b\end{pmatrix} 
+ \begin{pmatrix}b & \hphantom{-}0 \cr 0 & -b\end{pmatrix} 
= \begin{pmatrix}2 b & -\Omega   \cr  \Omega & \hphantom{-}0\end{pmatrix}, 
$$
which has
$$
\tr M_1 =2b, \quad \det M_1= \Omega^2 > 0,
$$
so this 2D system is stable for $b < 0$.

On the eigenspace with $\nu = -\overline \eta$ we get
$$
\begin{aligned}
\dot \eta  &= \left(  {K\over 6} \left(1\!-\!b^2\right)+i\left( 1\! -\! {2Kb \over 3}  \right) \right ) \eta + {K \over 6} \left( b^2\!-\!1\! +\!2ib  \right)\overline \eta 
\\
&= \left( b\Omega+ i(2 \Omega -1)  \right ) \eta + \left(-b \Omega+ i(1- \Omega) \right)  \overline \eta.
\end{aligned}
$$
The matrix for this 2D system is 
$$
M_{-1}  = \begin{pmatrix}b \Omega & 1- 2\Omega  \cr  2\Omega -1  & b \Omega\end{pmatrix} 
+ \begin{pmatrix}-b \Omega  & 1-\Omega  \cr 1 - \Omega   & b \Omega\end{pmatrix} 
= \begin{pmatrix}0 & 2-3\Omega   \cr  \Omega & 2b \Omega\end{pmatrix}, 
$$
which has
$$
\tr M_{-1}  =2b\Omega, \quad \det M_{-1} = (3\Omega-2 ) \Omega = {(1-5b^2)(1-b^2) \over (1+b^2)^2},
$$
so this 2D system is stable for $-\sqrt5/5 < b < 0$.  Combining both results, we see that the fixed point with $y_{\pm 1} = \pm ib$ is stable for 
(\ref{ReducedThreesys})\
provided that $-\sqrt5/5 < b < 0$.

In the original system 
(\ref{Threesys})\
we can re-write the $\dot \zeta$ equation as
$$
\dot \zeta = {K \over 2} \left(Y - \overline Y \right) \zeta,
$$
and we have $Y(t) - 1/3 \to 0$ decaying exponentially,  for initial conditions sufficiently near the fixed point $(-ib, 1, ib)$, provided that $-\sqrt5/5 < b < 0$.  This implies that $\zeta(t)$ will converge to a constant, and hence the order parameter $Z(t)$ converges to a constant of the form $\zeta_0/3$ with $\zeta_0 \in S^1$.

Now let's consider any trajectory $\gamma(t)$  in the double Poisson space $X_{DP}$ for the finite-$N$ continuum system with $N = 3$, $j = \omega_j = -1, 0,  1$.  Suppose the order parameter $Z(t) \to \zeta_0/3$ for some $\zeta_0 \in S^1$; by rotating $\gamma$ by $\overline \zeta_0$, we can assume that $Z(t) \to 1/3$.  So $\gamma$ has the same steady-state order parameter dynamics as the fixed points analyzed above. Any point  $(\rho_{-1}, \rho_0, \rho_1)$ in the forward limit set $L(\gamma)$  (which is nonempty since $X_{DP}$ is compact) must have $Z = 1/3$.  Represent any point in $L(\gamma)$  by a vector $(z_{-1}^{(1)} , z_{-1}^{(2)}, z_0^{(1)}, z_0^{(2)}, z_1^{(1)}, z_1^{(2)})$ with $z_j^{(\alpha)} \in \overline \Delta$.  Since $Z = 1/3$ is constant on $L(\gamma)$, the $z_j^{(\alpha)}$ must satisfy the equations
\begin{equation}
\label{Threesysdouble}
\begin{aligned}
\dot z_1^{(\alpha)} &= iz_1 ^{(\alpha)}+ {K \over 6} \left( 1 -  (z_1^{(\alpha)})^2\right ) \cr
\dot z_0^{(\alpha)} &=  {K \over 6} \left( 1 -  (z_0^{(\alpha)})^2\right ) \cr
\dot z_{-1} ^{(\alpha)} &= -iz_{-1}^{(\alpha)} + {K \over 6} \left( 1 -  (z_{-1}^{(\alpha)})^2 \right), \quad \alpha = 1, 2.
\end{aligned}
\end{equation}

We can solve the equations in 
(\ref{Threesysdouble})\
explicitly; for example the first equation (momentarily dropping the sub- and superscripts) is equivalent to
$$
-{6 \over K} {dz \over dt} = z^2 -{6 i z \over K} - 1 = (z-ib) (z-i b^{-1}),
$$
which can be integrated via partial fractions to obtain
$$
{z - ib \over z - i b^{-1}} = e^{i \Omega t} {z(0) - ib \over z(0) - i b^{-1}}.
$$
Let $H_1$ be the M\"obius map given by
$$
H_1(z) = {z-ib \over z-ib^{-1}};
$$
then
$$
z = ib^{-1} + {i(b - b^{-1} ) \over 1-e^{i \Omega t} H_1(z(0))}.
$$
We have $|b| < 1$, which implies that $|H_1(z)| < 1$ for any $z \in \overline \Delta$; therefore we can expand
$$
z = ib +i(b-b^{-1}) \sum_{n=1}^\infty e^{i n \Omega   t} H_1(z(0) ) ^n.
$$
Similarly, the solution to the third equation in 
(\ref{Threesysdouble})\
is
$$
z = -ib - i(b-b^{-1}) \sum_{n=1}^\infty e^{-i n\Omega   t} H_{-1}(z(0) ) ^n, 
$$
with
$$
H_{-1}(z) = {z+ib \over z+ib^{-1}}.
$$
The middle equation with $\omega_0 = 0$ has fixed points at $z = \pm1$; for $K < 0$ we have $z=-1$ attracting and $z=1$ repelling.  Any solution $z(t)$ to this equation will converge to $-1$, unless it is the fixed point $z(t) = 1$.  
 
Therefore any solution to 
(\ref{Threesysdouble})\
will have order parameter
\begin{widetext}
$$
\begin{aligned}
Z(t)  &=  {z_0^{(1)} (t) + z_0^{(2)} (t) \over 6}    + 
 { {i(b-b^{-1}) \over 6}}    \sum_{n=1}^\infty
\left[
  e^{i n \Omega   t} \Bigl (H_1(z_1^{(1)}(0) ) ^n \! + \! H_1(z_1^{(2)}(0))^n \Bigr )
 - 
e^{-i n \Omega   t} \Bigl (H_{-1}(z_{-1}^{(1)}(0) ) ^n \! + \! H_{-1}(z_{-1}^{(2)}(0)) ^n \Bigr).    
\right]
\end{aligned}
$$
\end{widetext}
Now we must have $Z(t) = 1/3$ for all $t$.  This can only occur if all the coefficients of $e^{\pm i n \Omega t}$ are $0$, and  the convergent terms $z_0^{(\alpha)}(t)$ are constant $= 1$.  Hence we must have
$$
H_1(z_1^{(1)}(0) ) + H_1(z_1^{(2)}(0) ) = H_1(z_1^{(1)}(0) )^2 + H_1(z_1^{(2)}(0) )^2 = 0,
$$
which implies $ H_1(z_1^{(1)}(0) ) = H_1(z_1^{(2)}(0) ) = 0$, and hence $ z_1^{(1)}(0)  = z_1^{(2)}(0)  = ib$; similarly $ z_{-1}^{(1)}(0)  = z_{-1}^{(2)}(0)  = -ib$.  So we have shown that the only trajectory in $X_{DP}$ for 
(\ref{Threesysdouble})\
which has $Z(t) = 1/3$ constantly is the fixed point $(-ib, -ib, 1, 1, ib, ib)$, which is the fixed point $(-ib, 1, ib)$  in $X_P$.
 
The argument above shows that if $Z(t) \to 1/3$ for some trajectory $\gamma(t)$ in $X_{DP}$, then the forward limit set of $\gamma$ must be the single point $(-ib,1,ib)$ in $X_{P}$; in particular, this implies $\gamma(t) \to (-ib,1,ib)$.  But if the initial point of $\gamma$ has either $z_1^{(1)} \ne z_1^{(2)}$ or $z_{-1}^{(1)} \ne z_{-1}^{(2)}$, then $\gamma(t)$ cannot converge to $(-ib, 1, ib)$, because the hyperbolic distances $d_{hyp} (z_{\pm 1}^{(1)}(t), z_{\pm 1}^{(2)}(t))$ are constant.  In other words, if we perturb the initial condition off the fixed point $(-ib, 1, ib)$ by splitting the Poisson densities with centroids $\pm ib$ into double Poissons, then the long-term behavior of the order parameter $Z(t)$ will not have the same steady-state dynamics $|Z(t)| \to 1/3$ that we get for perturbations of the fixed point inside the Poisson manifold
(see Figure~4).

\begin{figure}[h]
	\centerline{\includegraphics[width=2.8in]{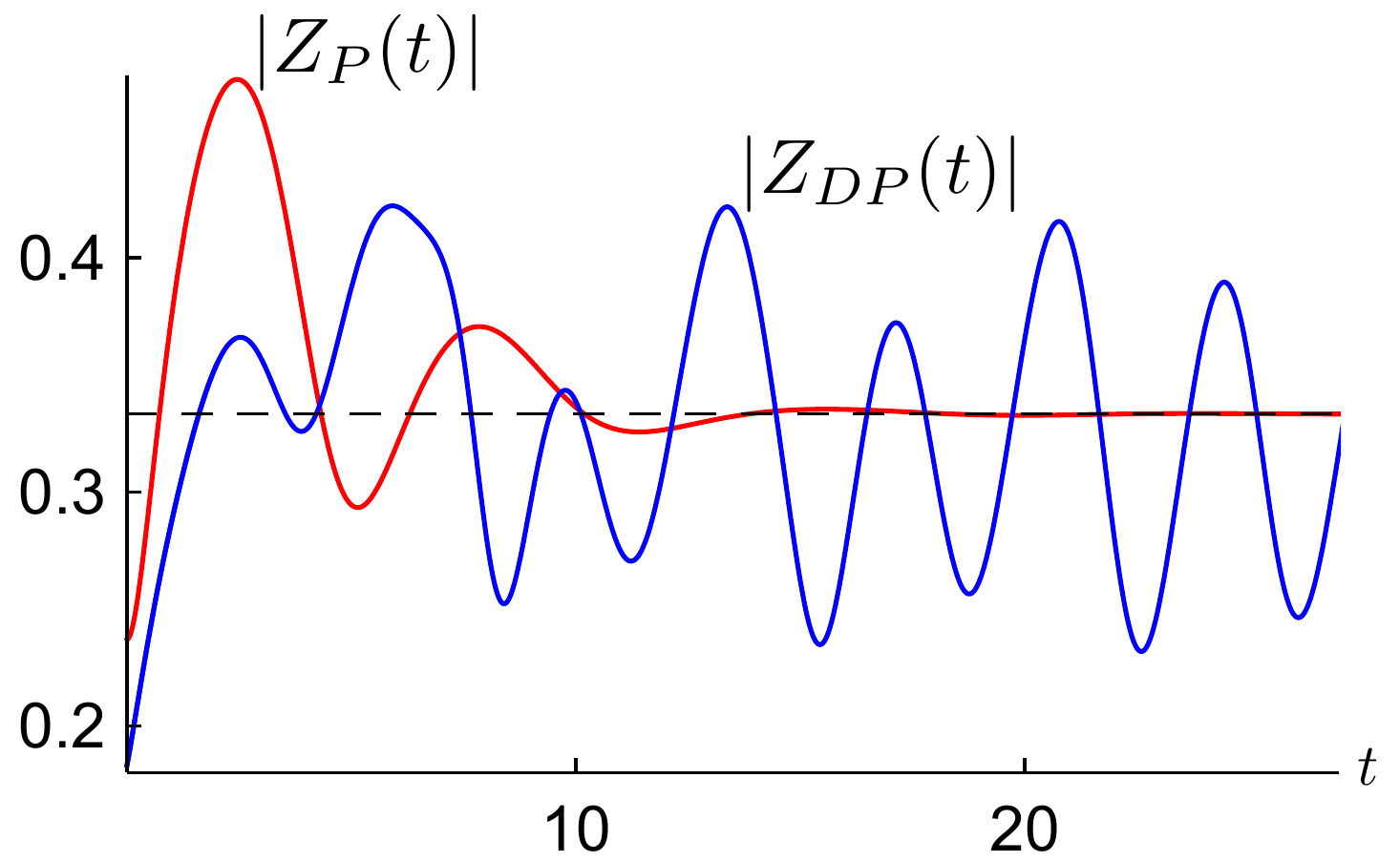}}
	\caption{\label{fig4}
For $-\sqrt{5}/5<b<0$ the example fixed point is stable on the Poisson manifold $X_P$; 
the order parameter $|Z_P(t)|\to\frac13$ (in red).
This fixed point is unstable 
off $X_{P}$;
the order parameter asymptotic dynamics  ($|Z_{DP}(t)|$  in blue)  are qualitatively different.
}
\end{figure}

\subsection{Multi-Poissons Are Dense}
 
\noindent
We conclude this section with a remark on the level of generality of averaged Poisson measures.  The Poisson measure $\rho$ with centroid $z \in \Delta$ has density function
$$
g_z(\zeta) = {1 - |z|^2 \over |\zeta - z|^2}.
$$
Let $f$ be a continuous function on $S^1$; then the classic Poisson integral of $f$ is the function on $\Delta$ defined by
$$
\tilde f(z) = {1 \over 2 \pi i}\int_{S^1} f(\zeta) 
\;
g_z(\zeta)
\;
{d \zeta \over \zeta}.
$$
It is well-known 
\cite{rudin2006real}
that $\tilde f$ is a continuous extension of $f$ to the disc $\Delta$; therefore the function $f_r$ on $S^1$ defined by
$f_r(\zeta) = \tilde f(r \zeta)$ for $0<r <1$ converges uniformly to $f$ on $S^1$ as $r \to 1$.  We have
$$
f_r(\zeta) = {1 \over 2 \pi } 
\int_0^{2 \pi} 
\!\!\!
f(e^{is} ) {1 - r^2 \over |e^{is}- r \zeta|^2} ds = {1 \over 2 \pi } 
\int_0^{2 \pi}
\!\!\! 
f(e^{is} ) {1 - r^2 \over |\zeta - r e^{is}|^2} \, ds. 
$$
 
Now suppose $f$ is a density function on $S^1$, so $f \ge 0$ and 
$$
{1 \over 2 \pi} \int_0^{2 \pi} f(e^{is} ) \, ds = 1.  
$$
Fix $r$, and consider the regular partition of $ [0, 2 \pi]$ with intervals $[s_{j-1}, s_j]$, $j = 1, \dots, n$ and $\Delta s = 2 \pi/n$.  For any $ \epsilon > 0$ we can choose $n$ large enough so that there exist $s_j^\ast \in [s_{j-1}, s_j]$ such that
$$
{1 \over 2 \pi} \sum_{j=1}^n f(s_j^\ast) \Delta s = 1
$$
and
$$
| f(s) - f(s_j^\ast) | < \epsilon,  \quad   \left |  {1 -r^2 \over |\zeta - r e^{is}|^2} -  {1-r^2  \over |\zeta - r e^{is_j^\ast}|^2} \right| < \epsilon
$$
for all $s \in [s_{j-1}, s_j]$.  This implies that as $\epsilon \to 0$ the sums
$$
{1 \over 2 \pi} \sum_{j=1} ^n f(e^{is_j^\ast} ) {1 - r^2 \over |\zeta - r e^{is_j^\ast }|^2} \Delta s  \to f_r(\zeta)
$$
uniformly in $\zeta$.  The sum above is a weighted average of Poisson density functions, 
with centroids $z_j=re^{is_j^\ast }$ and weights $w_j=f(s_j^\ast)\Delta s$.
This shows that
we can uniformly approximate $f_r$, and hence any continuous density function $f$, by weighted averages of Poisson densities.  
 
The above argument shows that the uniform closure of the set of weighted averages of Poisson measures is the set of all probability measures with continuous density functions on $S^1$.  In the natural topology on $Pr(S^1)$, which is weaker than the uniform topology on continuous densities, measures with continuous density functions are dense, so in the natural topology the set of weighted averages of Poisson measures is dense in $Pr(S^1)$.  Consequently, in principle all of the dynamics for the finite-$N$ continuum system will be revealed on the subset of weighted averages of Poisson measures.

\section{Infinite--N Continuum System}

\subsection{System Set-up}

\noindent
Next, we turn to the infinite--$N$ continuum version of 
(\ref{Ksys}), 
which is the setting for the famous Ott-Antonsen ansatz and analysis.  In this model we consider the frequency $\omega \in \Bbb R$ to vary continuously, according to a density function $g(\omega)$, which we will take to be the Lorentzian density function
$$
g(\omega) = {1 \over \pi} \cdot {1 \over \omega^2 + 1}.
$$
A state of the system consists of a family of probability measures $\omega \mapsto \rho_\omega$ parametrized by $\omega \in \Bbb R$; in other words, a state is a function $f : {\Bbb R} \to Pr(S^1)$.  We need at least a mild regularity condition on the function $f$; in 
\cite{mirollo2007spectrum}
we assumed only that the map $f$ is measurable.  Let $X$ be the state space consisting of all measurable families $\rho_\omega$.  For any $\rho_\omega \in X$, we can define the order parameter
$$
Z = \int_{\Bbb R} \int_{S^1} \zeta \, d \rho_\omega (\zeta)  g(\omega) \, d \omega,
$$
which naturally generalizes 
(\ref{ZF})\ 
to the infinite--$N$ continuum case. The evolution equation for the state $\rho_\omega$ is
\begin{equation}
\label{KsysINFC}
\dot \rho_\omega + {\partial \over \partial \zeta} \left( v_\omega \rho_\omega \right) = 0,
\quad {\rm with} \quad
v_\omega (\zeta) =  i \omega \zeta + {K \over 2} \left(Z - \overline Z \zeta^2 \right). 
\end{equation}
As in the finite--$N$ continuum case, for each $\omega$ the measure $\rho_\omega$ evolves in its M\"obius group orbit $G \rho_\omega$.  

\subsection{Poisson and OA Manifolds}

\noindent
As in the finite--$N$ continuum case, the Poisson manifold $X_P$ consisting of Poisson densities $\rho_\omega$ for each $\omega$ is an invariant subspace under the dynamics.  Since a Poisson measure is determined by its centroid $z \in \overline \Delta$, states in the Poisson manifold are determined by (measurable) functions $f: {\Bbb R} \to \overline \Delta$, $z_\omega = f(\omega)$.  The dynamics on the Poisson manifold are given by the system
\begin{equation}
\label{IPCsys}
\dot z_\omega  =  i \omega z_\omega + {K \over 2} \left( Z - \overline Z z_\omega^2 \right), 
\end{equation}
with
$$
Z = \int_{\Bbb R} z_\omega g(\omega) \, d \omega.
$$

The key to the ingenious calculation in Ott and Antonsen's famous paper 
\cite{ott2008low}
is to assume a rather strong regularity condition on $f$, namely that $f$ extends to an analytic function in the upper half plane $\Re \omega > 0$, which is bounded and approaches $0$ as $|\omega| \to \infty$.  Ott and Antonsen proved that this condition is preserved by the system dynamics, so the OA manifold $X_{OA}$ consisting of Poisson states $f$ satisfying this additional condition is invariant.  (Actually, Ott and Antonsen parametrized their Poisson densities by the conjugate of the centroid, so their analytic continuation was in the lower half plane.)  The analysis of the order parameter $Z$ on $X_{OA}$ is facilitated by the analytic continuation condition:  as shown in 
\cite{ott2008low},
if we integrate 
(\ref{IPCsys})\ 
over $\Bbb R$ against $g(\omega)$ we obtain
$$
\dot Z = i \int_{\Bbb R} \omega f(\omega)  g(\omega) \, d \omega + {K \over 2} \left ( Z - \overline Z \int_{\Bbb R}  f(\omega)^2 g(\omega) \, d \omega \right).
$$
We express
$$
g(\omega) = {1 \over 2 \pi i} \left ( {1 \over \omega - i} - {1 \over \omega + i} \right)
$$
and use this to evaluate $Z$ and the other two integrals above by the method of residues: the functions in each integrand have a single pole at $\omega = i$ in the upper half plane, and converge to $0$ as $| \omega| \to \infty$.  Therefore
$$
\begin{aligned}
\int_{\Bbb R}  f(\omega) \, d \omega &=  f(i) = Z \\
\int_{\Bbb R} \omega f(\omega) \, d \omega &= i f(i) = iZ \\
\int_{\Bbb R}  f(\omega)^2 \, d \omega &=f(i)^2 = Z^2
\end{aligned}
$$
and we obtain the Ott-Antonsen evolution equation for $Z$ on $X_{OA}$:
\begin{equation}
\label{ZOA}
\dot Z = -Z +{K \over 2} \left( 1-|Z|^2 \right) Z = \left( {K \over 2} - 1\right) Z -{K\over 2} |Z|^2 Z. 
\end{equation}

The beauty of this equation is that it is independent of the details of the individual Poisson measures parametrized by $z_\omega$, and is also very easy to analyze:  the flow 
(\ref{ZOA})\ 
is radial, the origin is stable for $K \le 2$, and loses stability for $K > 2$ where a stable solution with $|Z| > 0$ exists.  Unfortunately, the analytic continuation condition really is necessary; the dynamics of $Z$ do not obey 
(\ref{ZOA})\ 
in the full Poisson manifold.  As shown in 
\cite{mirollo2012asymptotic}, 
there are initial conditions in $X_P$  for which $Z(t)$ does not decay to $0$ exponentially, as predicted by 
(\ref{ZOA}), 
when $K < 2$.

\subsection{Multi-Poisson and OA Manifolds: Dynamics off $X_P$ and $X_{OA}$}

\noindent
Now we address the main question: is the OA manifold attracting?  Analogous to the finite--$N$ continuum case, we can define the double Poisson manifold $X_{DP}$ and double OA manifold $X_{DOA}$.  The measures $\rho_\omega$ in $X_{DP}$ are averages of two Poisson measures, so can be parametrized by two functions  $f^{(1)} , f^{(2)} : {\Bbb R} \to \overline \Delta$ defining the centroids
$$
z_\omega^{(1)} = f^{(1)}(\omega), \quad z_\omega^{(2)} = f^{(2)}(\omega);
$$
for $X_{DOA}$ we assume these functions also satisfy the OA analytic continuation condition.   The $z_\omega^{(\alpha)}$ evolve according to the system
\begin{equation}
\label{DIPCsys}
\begin{aligned}
\dot z_\omega^{(\alpha)}  &=  i \omega z_\omega^{(\alpha)} + {K \over 2} \left( Z - \overline Z (z_\omega^{(\alpha)})^2 \right),  \quad \alpha = 1,2 \cr
Z &= {1 \over 2} \left ( Z^{(1)}+ Z^{(2)} \right), \quad  Z^{(\alpha)} =  \int_{\Bbb R} z_\omega^{(\alpha)} g(\omega) \, d \omega.
\end{aligned}
\end{equation}

If the functions satisfy the condition $|f^{(\alpha)} (\omega)| < 1$ for all $\omega$, then for each $\omega$  the hyperbolic distance
$d_{hyp} (  z_\omega^{(1)} ,  z_\omega^{(2)} )$ is invariant under the system dynamics.
Consider any state $f\in X_P$ such that 
the set
$
A=\{ \omega : |f(\omega)|<1   \}
$
has measure $>0$.
Suppose
 $t\mapsto(f_t^{(1)},f_t^{(2)})$
is a trajectory for (\ref{DIPCsys}) in $X_{DP}$
that converges to $f$ in some topology.
For any reasonably strong topology, say the $L^p$ topology with $1\le p\le\infty$,
this implies the existence of a subsequence
$t_n\to\infty$ such that 
$
f_{t_n}^{(1)},f_{t_n}^{(2)}
\to f
$
a.e.~\cite{rudin2006real}.
Therefore  
the distance
$d_{hyp} ( f_{t_n}^{(1)}(\omega),f_{t_n}^{(2)}(\omega))\to0  $
a.e.~on $A$.
But this  distance is invariant under the dynamics, so we must have 
$f_{0}^{(1)}=f_{0}^{(2)}$ a.e.~on $A$.
This proves that a trajectory in $X_{DP}$ with
$f_{0}^{(1)}\ne f_{0}^{(2)}$ on $A$
cannot converge to the state $f\in\ X_P$.

Hence
for any reasonably strong topology on $X_{DP}$, 
{\sl the Poisson manifold $X_P$ is not attracting} for the dynamics given by 
(\ref{DIPCsys}): 
it is impossible to approach a state $\rho_\omega$ with $|z_\omega| < 1$ from $X_{DP} - X_P$.
The same is true for $X_{OA}$ inside $X_{DOA}$.  This is all perfectly analogous to  the finite--$N$ continuum case.  

\subsection{Order Parameter Dynamics off $X_{OA}$}

\noindent
If we restrict our attention just to the macroscopic order parameter $Z$, as is often the practice in studying infinite-$N$ systems, then something different happens compared to the finite--$N$ continuum case.
Consider any state  $(f^{(1)} , f^{(2)} )$ in $X_{DOA}$  with corresponding order parameters $Z^{(1)},  Z^{(2)}$.  A residue calculation exactly as  in the single Poisson case gives the equations
$$
\dot Z^{(\alpha)} = - Z^{(\alpha)} + {K \over 2} \left ( Z - \overline Z (Z^{(\alpha)})^2 \right), \quad \alpha = 1, 2
$$
We will show that these equations imply that $| Z^{(1)} -  Z^{(2)}| \to 0$ decaying exponentially, which implies that the dynamics of the average $Z$ are the same as on the manifold $X_{OA}$.  In other words, the long-term order parameter dynamics on $X_{DOA}$ and $X_{OA}$ are the same.  The crucial ingredient here is the  $- Z^{(\alpha)}$ term coming from the residue calculation.  To see this, consider any flow on $\overline\Delta$ of the form
\begin{equation}
\label{STAR}
\dot z = -z + Y - \overline Y z^2, 
\end{equation}
where $Y$ can depend on time $t$, but we assume $|Y|$ is bounded.  If $z(t)$ satisfies 
(\ref{STAR}),
then
$$
\begin{aligned}
(|z|^2)\, \dot {}& = \dot z \overline z + z \dot{\overline z} =  (-z + Y - \overline Y z^2 ) \overline z + z ( -\overline z + \overline Y - Y \overline  z^2) \cr
&= -2|z|^2 +2 (1 - |z|^2) \Re( \overline Y z).
\end{aligned}
$$
Observe that $(|z|^2)\, \dot {} = -2$ if $|z| = 1$; since $|Y|$ is assumed bounded, we can find $0 < r < 1$ so that $(|z|^2)\, \dot {} \le -1$ on the annulus $r \le |z| \le 1$.  Therefore any solution $z(t)$ must have $|z(t) |< r$ for $t$ sufficiently large (actually for $t \ge 1-r^2$).

Suppose we have two solutions $z(t), w(t)$ to an equation of the form 
(\ref{STAR});
we wish to prove that the hyperbolic distance $d_{hyp}(z,w) \to 0$ decaying exponentially.  The hyperbolic metric $d_{hyp}(z,w)$ is given by equations 
(\ref{HYPONE})\
and 
(\ref{HYPTWO}).
We claim that $\delta(z,w)$ satisfies the equation
$$
\dot \delta  = - \delta \Re \left ( { 1+z\overline w \over 1- z \overline w} \right) ;
$$
we can derive this directly using 
(\ref{STAR}),
though there is a better way that avoids this tedious calculation.  Observe that $Y - \overline Y z^2$ is an infinitesimal isometry for the hyperbolic geometry on the disc, and therefore will have no affect on the conformal invariant $\delta$; in other words, one can assume that $Y = 0$, $\dot z = -z$, $\dot w = -w$  and get the general result.  (If this seems like magic, we assure the skeptical reader that we carefully performed this calculation including the $Y$ terms, and saw that indeed they drop out.  It was only afterwards that we realized why this had to happen.)  The quantity $\lambda(z,w)$ from 
(\ref{HYPONE})\
has
$$
\dot \lambda = { (1- z\overline w  ) (w-z) -(z-w) \cdot 2 z\overline w   \over (1-z \overline w )^2} = { (w-z) (1+ z \overline w) \over (1 - z \overline w)^2};
$$
$\delta^2  = \lambda \overline \lambda$,  so
$$
\begin{aligned}
\delta \dot \delta &= \Re (\dot \lambda \overline \lambda)  = \Re \left(  { (w-z) (1+ z \overline w) \over (1 - z \overline w)^2} {\overline z - \overline w \over 1-\overline z w} \right) 
\\&= -\delta^2 \Re  \left ( { 1+z\overline w \over 1- z \overline w} \right),
\end{aligned}
$$
which gives the desired result.  

Next, observe that
$$ 
\begin{aligned}
 \Re  \left ( { 1+z\overline w \over 1- z \overline w} \right) &={ \Re \left (  (1+z\overline w) (1 - \overline z w) \right )\over |1- z \overline w|^2 } = {1 - |zw|^2 \over |1-z \overline w|^2} 
 \\&
 \ge {1-r^2 \over 1+r^2} = c >0
 \end{aligned}
 $$
for $|z|, |w| \le r$.  Any two solutions $z(t), w(t)$ to 
(\ref{STAR})\
will satisfy $|z|, |w| < r$ after time $t_0 =1-r^2$, and then $\delta(z,w)$ will decay exponentially, dominated by $e^{-ct}$; in other words, we have proved that
$$
\delta(t) \le e^{-c(t-t_0) }\delta (t_0)
$$
and therefore
$$
\begin{aligned}
&d_{hyp} (z(t), w(t) ) = 2 \left( \delta(t)  + {\delta(t)^3 \over 3} + {\delta(t) ^5 \over 5} + \cdots \right) \cr
&\le 2e^{-c(t-t_0) }\left ( \delta(t_0)  + { e^{-2c(t-t_0) }\delta(t_0)^3 \over 3} + {e^{-4c(t-t_0) }\delta(t) ^5 \over 5} + \cdots \right) \cr
&\le e^{-c(t-t_0) } d_{hyp} (z(t_0), w(t_0) )
\end{aligned}
$$
for $t \ge t_0$.  We also see that the Euclidean distance $|z(t) - w(t) | \to 0$ dominated by $e^{-ct}$, since the distortion factor $|z-w| / d_{hyp} (z,w)$ is bounded  if $|z| , |w| \le r < 1$.  This completes the proof that $| Z^{(1)} -  Z^{(2)}| \to 0$, decaying exponentially.  A slight variation of this argument extends to the case of weighted averages of Poissons on the analogous generalized OA manifold.

We conclude this section with an explanation of how the OA or Poisson manifold could still in some sense be attracting, in light of the discussion above.  A state consisting of double Poissons, with centroids $z^{(1)}_\omega \ne z^{(2)}_\omega$, can converge under the dynamics  in a weak sense to the Poisson manifold.  For example, consider the system 
(\ref{DIPCsys})\ 
with coupling $K = 0$; then each $z^{(\alpha)}_\omega$ evolves independently, giving
$$
z^{(\alpha)}_\omega (t) = e^{i \omega t} z^{(\alpha)}_\omega (0), \quad \alpha = 1,2.
$$
The functions $z^{(\alpha)}_\omega (t)$ converge to $0$ weakly as $t \to \infty$, as a consequence of the Riemann-Lebesgue lemma\cite{rudin2006real}: for any integrable function $f$ on $\Bbb R$, we have
$$
\int_{\Bbb R}  e^{i \omega t} z^{(\alpha)}_\omega (0) f(\omega) \, d \omega \to 0
$$
as $t \to \infty$.  
Note that this conclusion does not depend on any analytic continuation assumptions on the initial functions $f^{(\alpha)}$.
Our arguments above show that weak convergence to the OA manifold is all that one could hope for; on the other hand,
weak convergence is all that is required to capture the order parameter dynamics.

\section{Conclusion}

\noindent
For the finite--$N$ continuum Kuramoto model, the Poisson manifold is generally not attracting, and does not capture the complexity of the dynamics on the full state space.  We demonstrated this by defining the larger double Poisson manifold, and showed that one can assign  a measure of the distance of a state to the Poisson manifold which is dynamically invariant.  We also gave explicit examples of how the dynamics can become more complicated off the Poisson manifold.  
This has important consequences for the study of this and more general finite--$N$ continuum models.  
For example, one can investigate  multi-population models which have different coupling within populations compared to across populations.
In this study researchers have focused on so-called chimera states, in which some of the populations are in sync, whereas others are distributed according to smooth Poisson measures.  
We plan to address the consequences of our methodology to this class of models in detail in a follow-up paper, 
so we offer here only some  brief remarks on this topic.
Our analysis above easily extends to this setting and implies that
chimera states are {\it not}  attracting in the full state space,
even if they are attracting within the OA manifold.

The simplest chimeras occur for the model with $N=2$ populations studied in
Ref.~\cite{abrams2008solvable},
which has a 4D OA manifold consisting of pairs of Poisson densities.
For this model, chimera states are fixed states consisting of one smooth and one delta-function Poisson;
depending on parameters,
chimeras may exist and be stable in the OA manifold.
But off the OA manifold, the dynamics is restricted to 6D group orbits.
Chimera states correspond to limit cycles inside the {\it augmented} OA manifold (consisting of marked Poisson densities) and
therefore we must have stable limit cycle dynamics on the group orbits sufficiently near the OA manifold, which do {\it not}
relax back to chimera states on the OA.
This $N=2$ model can also have stable limit cycles {\it within} the OA manifold called breathing chimeras; 
these cycles correspond to stable quasi-periodic orbits in the augmented OA manifold.
Therefore, perturbing off the OA will also result in stable quasi-periodic dynamics;
which again do not relax back to the breathing chimeras on the OA.
Our methodology rigorously establishes
these  dynamic phenomena, 
which were conjectured 
and supported numerically in ref \cite{pikovsky2008partially}.
Our analysis also suggests a potential pitfall in numerical simulations of finite-$N$ continuum models.
If one approximates a continuum chimera state (stable in the OA) with $M$ discrete oscillators approximating the chimera distribution,
then this discrete oscillator population will have limit cycle dynamics on its group orbit for sufficiently large $M$, and cannot flow to sync.
However, numerical simulations may fail to reveal periodic dynamics because
the numerically approximated trajectories will not generally be constrained to remain on the original group orbit.

Perhaps most importantly,
our method using double Poissons or more general weighted averages of Poissons provides a framework for systematically exploring the dynamics of multi-population finite--$N$ continuum models off the Poisson manifold,  which can
reveal the more complex dynamics that is missed by focusing exclusively on Poisson states.
The story is more subtle for the infinite--$N$ continuum Kuramoto model, and comes down to a matter of interpretation as to what ``attracting'' really means.  The Poisson and OA manifolds are not attracting in the traditional sense, meaning trajectories starting sufficiently close to these manifolds converge to them in some reasonable topology.  We demonstrated this by defining similar measures of distance to the Poisson or OA manifolds on the larger double Poisson versions of these manifolds, which are again dynamically invariant.  So on the level of individual measures, we don't get convergence to the Poisson or OA manifolds.  However, the sense in which the OA manifold may be considered attracting is that 
with appropriate assumptions of the initial state,
the macroscopic order parameter $Z$ for the system has the same steady-state dynamics on the larger double Poisson version of these manifolds, or more generally weighted average Poisson versions.
Essentially, going to multiple Poisson densities does not effect the macroscopic order parameter dynamics, as long as
the functions parametrizing the families of Poisson measures satisfy the OA analyticity conditions.
We explicitly demonstrated this using  hyperbolic geometry techniques.

To sum up, the technique of multiple Poisson manifolds provides a systematic framework for studying the dynamics of multi-population
continuum Kuramoto networks beyond the restrictions of the Poisson and OA manifolds, and has the potential to reveal more intricate
dynamical behavior than has  previously been observed for these networks.
 
 We wish to thank Martin Bridgeman and Steven Strogatz for many helpful conversations in the course of preparing this manuscript.  This research was supported by NSF DMS 1413020.

\bibliography{refs_Jan2020}

\begin{thebibliography}{21}%
\makeatletter
\providecommand \@ifxundefined [1]{%
 \@ifx{#1\undefined}
}%
\providecommand \@ifnum [1]{%
 \ifnum #1\expandafter \@firstoftwo
 \else \expandafter \@secondoftwo
 \fi
}%
\providecommand \@ifx [1]{%
 \ifx #1\expandafter \@firstoftwo
 \else \expandafter \@secondoftwo
 \fi
}%
\providecommand \natexlab [1]{#1}%
\providecommand \enquote  [1]{``#1''}%
\providecommand \bibnamefont  [1]{#1}%
\providecommand \bibfnamefont [1]{#1}%
\providecommand \citenamefont [1]{#1}%
\providecommand \href@noop [0]{\@secondoftwo}%
\providecommand \href [0]{\begingroup \@sanitize@url \@href}%
\providecommand \@href[1]{\@@startlink{#1}\@@href}%
\providecommand \@@href[1]{\endgroup#1\@@endlink}%
\providecommand \@sanitize@url [0]{\catcode `\\12\catcode `\$12\catcode
  `\&12\catcode `\#12\catcode `\^12\catcode `\_12\catcode `\%12\relax}%
\providecommand \@@startlink[1]{}%
\providecommand \@@endlink[0]{}%
\providecommand \url  [0]{\begingroup\@sanitize@url \@url }%
\providecommand \@url [1]{\endgroup\@href {#1}{\urlprefix }}%
\providecommand \urlprefix  [0]{URL }%
\providecommand \Eprint [0]{\href }%
\providecommand \doibase [0]{http://dx.doi.org/}%
\providecommand \selectlanguage [0]{\@gobble}%
\providecommand \bibinfo  [0]{\@secondoftwo}%
\providecommand \bibfield  [0]{\@secondoftwo}%
\providecommand \translation [1]{[#1]}%
\providecommand \BibitemOpen [0]{}%
\providecommand \bibitemStop [0]{}%
\providecommand \bibitemNoStop [0]{.\EOS\space}%
\providecommand \EOS [0]{\spacefactor3000\relax}%
\providecommand \BibitemShut  [1]{\csname bibitem#1\endcsname}%
\let\auto@bib@innerbib\@empty
\bibitem [{\citenamefont {Kuramoto}(1975)}]{kuramoto1975self}%
  \BibitemOpen
  \bibfield  {author} {\bibinfo {author} {\bibfnamefont {Y.}~\bibnamefont
  {Kuramoto}},\ }in\ \href@noop {} {\emph {\bibinfo {booktitle} {International
  symposium on mathematical problems in theoretical physics}}}\ (\bibinfo
  {organization} {Springer},\ \bibinfo {year} {1975})\ pp.\ \bibinfo {pages}
  {420--422}\BibitemShut {NoStop}%
\bibitem [{\citenamefont {Strogatz}\ and\ \citenamefont
  {Mirollo}(1991)}]{strogatz1991stability}%
  \BibitemOpen
  \bibfield  {author} {\bibinfo {author} {\bibfnamefont {S.~H.}\ \bibnamefont
  {Strogatz}}\ and\ \bibinfo {author} {\bibfnamefont {R.~E.}\ \bibnamefont
  {Mirollo}},\ }\href@noop {} {\bibfield  {journal} {\bibinfo  {journal}
  {Journal of Statistical Physics}\ }\textbf {\bibinfo {volume} {63}},\
  \bibinfo {pages} {613} (\bibinfo {year} {1991})}\BibitemShut {NoStop}%
\bibitem [{\citenamefont {Strogatz}(2000)}]{strogatz2000kuramoto}%
  \BibitemOpen
  \bibfield  {author} {\bibinfo {author} {\bibfnamefont {S.~H.}\ \bibnamefont
  {Strogatz}},\ }\href@noop {} {\bibfield  {journal} {\bibinfo  {journal}
  {Physica D: Nonlinear Phenomena}\ }\textbf {\bibinfo {volume} {143}},\
  \bibinfo {pages} {1} (\bibinfo {year} {2000})}\BibitemShut {NoStop}%
\bibitem [{\citenamefont {Pikovsky}\ and\ \citenamefont
  {Rosenblum}(2008)}]{pikovsky2008partially}%
  \BibitemOpen
  \bibfield  {author} {\bibinfo {author} {\bibfnamefont {A.}~\bibnamefont
  {Pikovsky}}\ and\ \bibinfo {author} {\bibfnamefont {M.}~\bibnamefont
  {Rosenblum}},\ }\href@noop {} {\bibfield  {journal} {\bibinfo  {journal}
  {Physical review letters}\ }\textbf {\bibinfo {volume} {101}},\ \bibinfo
  {pages} {264103} (\bibinfo {year} {2008})}\BibitemShut {NoStop}%
\bibitem [{\citenamefont {Ott}\ and\ \citenamefont
  {Antonsen}(2008)}]{ott2008low}%
  \BibitemOpen
  \bibfield  {author} {\bibinfo {author} {\bibfnamefont {E.}~\bibnamefont
  {Ott}}\ and\ \bibinfo {author} {\bibfnamefont {T.~M.}\ \bibnamefont
  {Antonsen}},\ }\href@noop {} {\bibfield  {journal} {\bibinfo  {journal}
  {Chaos: An Interdisciplinary Journal of Nonlinear Science}\ }\textbf
  {\bibinfo {volume} {18}},\ \bibinfo {pages} {037113} (\bibinfo {year}
  {2008})}\BibitemShut {NoStop}%
\bibitem [{\citenamefont {Bick}\ \emph {et~al.}(2019)\citenamefont {Bick},
  \citenamefont {Goodfellow}, \citenamefont {Laing},\ and\ \citenamefont
  {Martens}}]{bick2019understanding}%
  \BibitemOpen
  \bibfield  {author} {\bibinfo {author} {\bibfnamefont {C.}~\bibnamefont
  {Bick}}, \bibinfo {author} {\bibfnamefont {M.}~\bibnamefont {Goodfellow}},
  \bibinfo {author} {\bibfnamefont {C.~R.}\ \bibnamefont {Laing}}, \ and\
  \bibinfo {author} {\bibfnamefont {E.~A.}\ \bibnamefont {Martens}},\
  }\href@noop {} {\bibfield  {journal} {\bibinfo  {journal} {arXiv preprint
  arXiv:1902.05307}\ } (\bibinfo {year} {2019})}\BibitemShut {NoStop}%
\bibitem [{\citenamefont {Abrams}\ and\ \citenamefont
  {Strogatz}(2004)}]{abrams2004chimera}%
  \BibitemOpen
  \bibfield  {author} {\bibinfo {author} {\bibfnamefont {D.~M.}\ \bibnamefont
  {Abrams}}\ and\ \bibinfo {author} {\bibfnamefont {S.~H.}\ \bibnamefont
  {Strogatz}},\ }\href@noop {} {\bibfield  {journal} {\bibinfo  {journal}
  {Physical review letters}\ }\textbf {\bibinfo {volume} {93}},\ \bibinfo
  {pages} {174102} (\bibinfo {year} {2004})}\BibitemShut {NoStop}%
\bibitem [{\citenamefont {Abrams}\ \emph {et~al.}(2008)\citenamefont {Abrams},
  \citenamefont {Mirollo}, \citenamefont {Strogatz},\ and\ \citenamefont
  {Wiley}}]{abrams2008solvable}%
  \BibitemOpen
  \bibfield  {author} {\bibinfo {author} {\bibfnamefont {D.~M.}\ \bibnamefont
  {Abrams}}, \bibinfo {author} {\bibfnamefont {R.}~\bibnamefont {Mirollo}},
  \bibinfo {author} {\bibfnamefont {S.~H.}\ \bibnamefont {Strogatz}}, \ and\
  \bibinfo {author} {\bibfnamefont {D.~A.}\ \bibnamefont {Wiley}},\ }\href@noop
  {} {\bibfield  {journal} {\bibinfo  {journal} {Physical review letters}\
  }\textbf {\bibinfo {volume} {101}},\ \bibinfo {pages} {084103} (\bibinfo
  {year} {2008})}\BibitemShut {NoStop}%
\bibitem [{\citenamefont {Laing}(2009)}]{laing2009chimera}%
  \BibitemOpen
  \bibfield  {author} {\bibinfo {author} {\bibfnamefont {C.~R.}\ \bibnamefont
  {Laing}},\ }\href@noop {} {\bibfield  {journal} {\bibinfo  {journal} {Chaos:
  An Interdisciplinary Journal of Nonlinear Science}\ }\textbf {\bibinfo
  {volume} {19}},\ \bibinfo {pages} {013113} (\bibinfo {year}
  {2009})}\BibitemShut {NoStop}%
\bibitem [{\citenamefont {Martens}\ \emph {et~al.}(2009)\citenamefont
  {Martens}, \citenamefont {Barreto}, \citenamefont {Strogatz}, \citenamefont
  {Ott}, \citenamefont {So},\ and\ \citenamefont
  {Antonsen}}]{martens2009exact}%
  \BibitemOpen
  \bibfield  {author} {\bibinfo {author} {\bibfnamefont {E.~A.}\ \bibnamefont
  {Martens}}, \bibinfo {author} {\bibfnamefont {E.}~\bibnamefont {Barreto}},
  \bibinfo {author} {\bibfnamefont {S.~H.}\ \bibnamefont {Strogatz}}, \bibinfo
  {author} {\bibfnamefont {E.}~\bibnamefont {Ott}}, \bibinfo {author}
  {\bibfnamefont {P.}~\bibnamefont {So}}, \ and\ \bibinfo {author}
  {\bibfnamefont {T.~M.}\ \bibnamefont {Antonsen}},\ }\href@noop {} {\bibfield
  {journal} {\bibinfo  {journal} {Physical Review E}\ }\textbf {\bibinfo
  {volume} {79}},\ \bibinfo {pages} {026204} (\bibinfo {year}
  {2009})}\BibitemShut {NoStop}%
\bibitem [{\citenamefont {Martens}\ \emph {et~al.}(2013)\citenamefont
  {Martens}, \citenamefont {Thutupalli}, \citenamefont {Fourri{\`e}re},\ and\
  \citenamefont {Hallatschek}}]{martens2013chimera}%
  \BibitemOpen
  \bibfield  {author} {\bibinfo {author} {\bibfnamefont {E.~A.}\ \bibnamefont
  {Martens}}, \bibinfo {author} {\bibfnamefont {S.}~\bibnamefont {Thutupalli}},
  \bibinfo {author} {\bibfnamefont {A.}~\bibnamefont {Fourri{\`e}re}}, \ and\
  \bibinfo {author} {\bibfnamefont {O.}~\bibnamefont {Hallatschek}},\
  }\href@noop {} {\bibfield  {journal} {\bibinfo  {journal} {Proceedings of the
  National Academy of Sciences}\ }\textbf {\bibinfo {volume} {110}},\ \bibinfo
  {pages} {10563} (\bibinfo {year} {2013})}\BibitemShut {NoStop}%
\bibitem [{\citenamefont {Chiba}\ and\ \citenamefont
  {Nishikawa}(2011)}]{chiba2011center}%
  \BibitemOpen
  \bibfield  {author} {\bibinfo {author} {\bibfnamefont {H.}~\bibnamefont
  {Chiba}}\ and\ \bibinfo {author} {\bibfnamefont {I.}~\bibnamefont
  {Nishikawa}},\ }\href@noop {} {\bibfield  {journal} {\bibinfo  {journal}
  {Chaos: An Interdisciplinary Journal of Nonlinear Science}\ }\textbf
  {\bibinfo {volume} {21}},\ \bibinfo {pages} {043103} (\bibinfo {year}
  {2011})}\BibitemShut {NoStop}%
\bibitem [{\citenamefont {Chiba}(2015)}]{chiba2015proof}%
  \BibitemOpen
  \bibfield  {author} {\bibinfo {author} {\bibfnamefont {H.}~\bibnamefont
  {Chiba}},\ }\href@noop {} {\bibfield  {journal} {\bibinfo  {journal} {Ergodic
  Theory and Dynamical Systems}\ }\textbf {\bibinfo {volume} {35}},\ \bibinfo
  {pages} {762} (\bibinfo {year} {2015})}\BibitemShut {NoStop}%
\bibitem [{\citenamefont {Ott}\ and\ \citenamefont
  {Antonsen}(2009)}]{ott2009long}%
  \BibitemOpen
  \bibfield  {author} {\bibinfo {author} {\bibfnamefont {E.}~\bibnamefont
  {Ott}}\ and\ \bibinfo {author} {\bibfnamefont {T.~M.}\ \bibnamefont
  {Antonsen}},\ }\href@noop {} {\bibfield  {journal} {\bibinfo  {journal}
  {Chaos: An interdisciplinary journal of nonlinear science}\ }\textbf
  {\bibinfo {volume} {19}},\ \bibinfo {pages} {023117} (\bibinfo {year}
  {2009})}\BibitemShut {NoStop}%
\bibitem [{\citenamefont {Mirollo}\ and\ \citenamefont
  {Strogatz}(2007)}]{mirollo2007spectrum}%
  \BibitemOpen
  \bibfield  {author} {\bibinfo {author} {\bibfnamefont {R.}~\bibnamefont
  {Mirollo}}\ and\ \bibinfo {author} {\bibfnamefont {S.~H.}\ \bibnamefont
  {Strogatz}},\ }\href@noop {} {\bibfield  {journal} {\bibinfo  {journal}
  {Journal of Nonlinear Science}\ }\textbf {\bibinfo {volume} {17}},\ \bibinfo
  {pages} {309} (\bibinfo {year} {2007})}\BibitemShut {NoStop}%
\bibitem [{\citenamefont {Chen}, \citenamefont {Engelbrecht},\ and\
  \citenamefont {Mirollo}(2017)}]{chen2017hyperbolic}%
  \BibitemOpen
  \bibfield  {author} {\bibinfo {author} {\bibfnamefont {B.}~\bibnamefont
  {Chen}}, \bibinfo {author} {\bibfnamefont {J.~R.}\ \bibnamefont
  {Engelbrecht}}, \ and\ \bibinfo {author} {\bibfnamefont {R.}~\bibnamefont
  {Mirollo}},\ }\href@noop {} {\bibfield  {journal} {\bibinfo  {journal}
  {Journal of Physics A: Mathematical and Theoretical}\ }\textbf {\bibinfo
  {volume} {50}},\ \bibinfo {pages} {355101} (\bibinfo {year}
  {2017})}\BibitemShut {NoStop}%
\bibitem [{\citenamefont {Ahlfors}(1973)}]{ahlfors1973conformal}%
  \BibitemOpen
  \bibfield  {author} {\bibinfo {author} {\bibfnamefont {L.~V.}\ \bibnamefont
  {Ahlfors}},\ }\href@noop {} {\emph {\bibinfo {title} {Conformal invariants:
  topics in geometric function theory}}}\ (\bibinfo  {publisher}
  {McGraw-Hill},\ \bibinfo {year} {1973})\BibitemShut {NoStop}%
\bibitem [{\citenamefont {Marvel}, \citenamefont {Mirollo},\ and\ \citenamefont
  {Strogatz}(2009)}]{marvel2009identical}%
  \BibitemOpen
  \bibfield  {author} {\bibinfo {author} {\bibfnamefont {S.~A.}\ \bibnamefont
  {Marvel}}, \bibinfo {author} {\bibfnamefont {R.~E.}\ \bibnamefont {Mirollo}},
  \ and\ \bibinfo {author} {\bibfnamefont {S.~H.}\ \bibnamefont {Strogatz}},\
  }\href@noop {} {\bibfield  {journal} {\bibinfo  {journal} {Chaos: An
  Interdisciplinary Journal of Nonlinear Science}\ }\textbf {\bibinfo {volume}
  {19}},\ \bibinfo {pages} {043104} (\bibinfo {year} {2009})}\BibitemShut
  {NoStop}%
\bibitem [{\citenamefont {Chen}, \citenamefont {Engelbrecht},\ and\
  \citenamefont {Mirollo}(2019)}]{chen2019dynamics}%
  \BibitemOpen
  \bibfield  {author} {\bibinfo {author} {\bibfnamefont {B.}~\bibnamefont
  {Chen}}, \bibinfo {author} {\bibfnamefont {J.~R.}\ \bibnamefont
  {Engelbrecht}}, \ and\ \bibinfo {author} {\bibfnamefont {R.}~\bibnamefont
  {Mirollo}},\ }\href@noop {} {\bibfield  {journal} {\bibinfo  {journal}
  {Chaos: An Interdisciplinary Journal of Nonlinear Science}\ }\textbf
  {\bibinfo {volume} {29}},\ \bibinfo {pages} {013126} (\bibinfo {year}
  {2019})}\BibitemShut {NoStop}%
\bibitem [{\citenamefont {Rudin}(1987)}]{rudin2006real}%
  \BibitemOpen
  \bibfield  {author} {\bibinfo {author} {\bibfnamefont {W.}~\bibnamefont
  {Rudin}},\ }\href@noop {} {\emph {\bibinfo {title} {Real and complex
  analysis}}}\ (\bibinfo  {publisher} {McGraw-Hill},\ \bibinfo {year}
  {1987})\BibitemShut {NoStop}%
\bibitem [{\citenamefont {Mirollo}(2012)}]{mirollo2012asymptotic}%
  \BibitemOpen
  \bibfield  {author} {\bibinfo {author} {\bibfnamefont {R.~E.}\ \bibnamefont
  {Mirollo}},\ }\href@noop {} {\bibfield  {journal} {\bibinfo  {journal}
  {Chaos: An Interdisciplinary Journal of Nonlinear Science}\ }\textbf
  {\bibinfo {volume} {22}},\ \bibinfo {pages} {043118} (\bibinfo {year}
  {2012})}\BibitemShut {NoStop}%
\end{thebibliography}%



\end{document}